\documentclass[aps,prl,twocolumn,groupedaddress,showpacs,amssymb,amsmath,letter]{revtex4}
\usepackage{dcolumn}
\usepackage{graphicx}

\begin{document}


\title{Hyperfine Interactions in Graphene and Related Carbon Nanostructures} 

\author{Oleg V. Yazyev}
\email[Electronic address: ]{oleg.yazyev@epfl.ch}
\altaffiliation{Now at: ITP-EPFL and IRRMA, CH-1015 Lausanne, Switzerland}	 
\affiliation{Ecole Polytechnique F\'ed\'erale de Lausanne (EPFL), Institute of Chemical Sciences and Engineering, CH-1015 Lausanne, Switzerland}      

\date{\today}


\begin{abstract}
 
Hyperfine interactions, magnetic interactions between the spins of electrons and nuclei, 
in graphene and related carbon nanostructures are studied. 
By using a combination of accurate first principles 
calculations on graphene fragments and statistical analysis, I show that
both isotropic and dipolar hyperfine interactions can be
accurately described in terms of the local electron spin distribution and atomic structure.
A complete set of parameters describing the hyperfine interactions of $^{13}$C
and other nuclear spins at substitution impurities and edge terminations
is determined. 

\end{abstract}

\maketitle

Graphene and related carbon nanostructures are considered as potential
building blocks of future electronics, including spintronics \cite{Zutic04} 
and quantum information processing based on electron spins \cite{Loss98} or nuclear 
spins \cite{Kane98}.
Carbon nanostructures are attractive for these applications because of 
the weak spin-orbit interaction in materials made of light elements 
\cite{Xiong04,Bader06}.
Promising results for the spin-polarized current lifetimes in 
carbon nanotubes \cite{Tsukagoshi99,Sahoo05b,Hueso07} 
and graphene \cite{Hill06} unambiguously confirm the potential of these materials. 
A number of quantum dot devices, components of solid-state quantum computers,
based on carbon nanostructures have been proposed 
recently \cite{Bockrath01,Buitelaar02,Trauzettel06,Silvestrov07}. 
Hyperfine interactions (HFIs), the
weak magnetic interactions between the spins of electrons and nuclei, 
become increasingly important on the nanoscale.
In carbon nanostructures the interactions of electron spins with 
an ensemble of nuclear spins are expected to be the leading contribution 
to the electron spin decoherence \cite{Xiong04,Sahoo05b,Semenov07}.
Minimizing HFIs is thus necessary for achieving longer 
electron spin coherence times \cite{Khaetskii02}. 
In some other instances the HFIs play an important role as a link between the spins of 
electrons and nuclei in the nanostructures \cite{Kane98,Taylor03,Epstein05,Childress06}
underlying the implementations of quantum information processing involving nuclear spins.
Probing HFIs with magnetic resonance techniques also provides a wealth of information about 
structure and dynamics of carbon materials \cite{Pennington96}. 
A common understanding and an ability to control the HFIs  
are thus necessary for engineering future electronic devices
based on graphene and related nanostructures.
 
In this Letter, I study the hyperfine interactions in carbon 
nanostructures by using a combination of accurate first principles 
calculations on graphene fragments and statistical analysis. I show that the interaction 
of the conduction (low-energy) $\pi$ electron spins 
with nuclear spins can be described in terms of only the local (on-site and first-nearest-neighbor) 
$\pi$ electron spin distribution and the local atomic structure. 
The conduction electron spin distribution can be 
determined using simpler computational approaches (e.g. tight binding or 
analytical approximations \cite{Nakada96,Pereira06})
and tuned by tailoring nanostructure dimensions and applying external fields \cite{Tifrea03,Poggio03}.
The local nature of HFIs justifies the extension 
of my results from small molecular models to extended systems.
I further extend the considerations to curved topologies and to the 
presence of heteronuclei at impurities and boundaries.

The all-electron density functional theory (DFT) calculations \cite{Gaussian03}
were performed using a combination of the EPR-III Gaussian orbital basis set \cite{Rega96}
specially tailored for the calculations of hyperfine couplings and 
the B3LYP exchange-correlation hybrid density functional \cite{Becke88-plus}. 
This computational protocol (see Ref.~\onlinecite{Hermosilla05} for details) can be applied to molecules 
of limited size and predicts hyperfine coupling constants (HFCCs) in excellent agreement with
experimental results \cite{Hermosilla05}. 
Spin-orbit and relativistic effects which are not important for the calculation of HFIs in light-element
systems \cite{vanLenthe98} have been neglected.
For a set of representative experimentally 
measured $^{13}$C isotropic HFCCs of graphenic ion-radicals 
\cite{{Bolton64-plus}}
our computations provide a mean absolute error of 1.1~MHz ($\approx$2\% 
of the range of magnitudes), which justifies the use of
calculated HFCCs as a reference. 

\begin{figure}
\includegraphics[width=8.5cm]{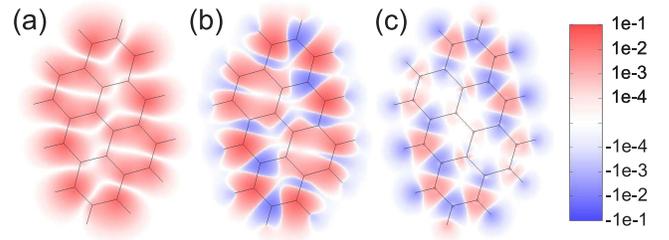}
\caption{\label{fig1}
(Color online). 
Projections of the spin-polarized conduction electron density $\rho^s_c(\mathbf{r})$ (a)
and the total spin density $\rho^s(\mathbf{r})$ (b) on the plane of an electron-doped graphene fragment (in a.u.$^{-2}$). 
The total spin density $\rho^s(\mathbf{r},z$$=$$0)$ (c) in 
the plane of nuclei (in a.u.$^{-3}$) reflects the isotropic hyperfine field. 
Molecular framework is shown by black lines. }  
\end{figure}

The effective spin-Hamiltonian of the HFI between the electron spin ${\mathbf S}$ and the nuclear 
spin ${\mathbf I}$ can be written as 
$\hat{\mathbf H}$=${\mathbf S}$$\cdot$$\overleftrightarrow{\mathbf  A}$$\cdot$${\mathbf I}$, where the
3$\times$3 HFI tensor 
$\overleftrightarrow{\mathbf  A}$$=$$A_{iso}$$\cdot$$\overleftrightarrow{\mathbf  1}$$+$$\overleftrightarrow{\mathbf T}$ 
is usually decomposed into the scalar 
HFCC $A_{iso}$ and the traceless dipolar HFI tensor 
$\overleftrightarrow{\mathbf T}$ \cite{Kaupp04}.  The HFI tensor $\overleftrightarrow{\mathbf A}$ reflects the distribution
of the electron spin density $\rho^s(\mathbf{r})$=$\rho^\uparrow(\mathbf{r})$$-$$\rho^\downarrow(\mathbf{r})$ 
viewed from the position of the nucleus $I$. 
In carbon nanostructures the nuclear spins are those of the $^{13}$C isotope 
($\approx$1.1\% natural abundance and can be atrificially changed; the dominant $^{12}$C isotope has zero spin) 
and other elements originating from impurities and boundaries.
The electron spin density $\rho^s({\mathbf r})$ can be further decomposed into the contribution of half-populated conduction electron states 
lying close to the Fermi level (or singly occupied molecular orbitals in the molecular context), $\rho^s_c({\mathbf r})$=$\sum_c |\psi^c({\mathbf r})|^2$$\ge$0, 
and the contribution of the fully populated valence states perturbed by the exchange
with spin-polarized conduction electrons, $\rho^s_{v}({\mathbf r})$=$\sum_v |\psi^{v\uparrow}({\mathbf r})|^2$$-$$|\psi^{v\downarrow}({\mathbf r})|^2$.
The crucial role of the exchange-polarization effect is illustrated with 
a model electron-doped hydrogen-terminated graphene fragment in the doublet
spin state (Fig.~\ref{fig1}). While the projection of $\rho^s_c({\mathbf r})$ on the $xy$ plane 
(Fig.~\ref{fig1}a) is positive everywhere and reveals an enhancement at the zig-zag edges, the projection of the total spin-density 
$\rho^s({\mathbf r})$ (Fig.~\ref{fig1}b)
is negative where $\rho^s_c({\mathbf r})$ is close to zero.
The isotropic (Fermi contact) HFCC is proportional to 
the total spin density at the position of nucleus $I$, 
$A_{iso}$$=$$(4\pi / 3 S) \beta_e\beta_N g_e g_I \rho^s({\mathbf r}_I)$, 
where $\beta_e$ and $\beta_N$ are the Bohr and nuclear magnetons, while $g_e$ and
$g_I$ are the $g$-values of free electron and nucleus $I$, respectively. $S$ is the maximum value
of the electron spin projection.
For the ideal graphene and planar $sp^2$ carbon nanostructures 
(all nuclei lie in the $z$=0 plane) $\rho^s_c(z$=$0)$=0 due to the $p_z$ symmetry of the 
conduction states.
However, there is a contribution of the $\sigma$
symmetry valence states $\rho^s_v(z$=$0)$$\neq$0 due to the exchange-polarization effect.
For the model graphene fragment $\rho^s_v(z$=$0)$ (Fig.~\ref{fig1}c) shows
an alternating pattern with a relative dominance of the negative
spin density.
Since the $\sigma$ states are situated well above and well below the Fermi level
in $sp^2$ carbon nanostructures, the valence exchange-polarization phenomenon 
exhibits the property of locality. This property was  
exploited by Karplus and Fraenkel almost 50 years ago to describe the isotropic
$^{13}$C HFCCs in conjugated organic radicals \cite{Karplus61}.
The main contribution to the hyperfine anisotropy
originates from the total spin population $n$ of the on-site $p_z$ atomic orbital, which
also incorporates the contribution of exchange-polarized valence states.
Assuming a local axial symmetry, $\overleftrightarrow{\mathbf T}$  can be 
written as a diagonal matrix with elements $T_{zz}$/2=$-T_{xx}$=$-T_{yy}$=$A_{dip}$, 
where $A_{dip}$=$(1/5 S) \beta_e\beta_N g_e g_I n\langle 1/r_{2p}^3\rangle$ ($r_{2p}$ is the distance of the carbon $2p$ electron to nucleus). 
      
\begin{figure}
\includegraphics[width=8.5cm]{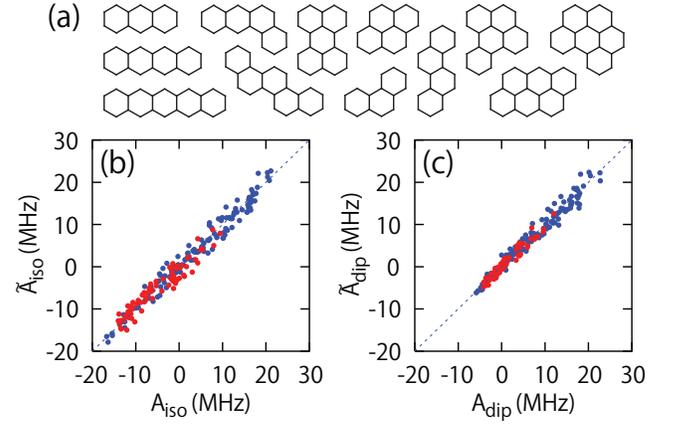}
\caption{\label{fig2} (Color online)
(a) Set of graphene fragments used in the present calculations.
Fitted Fermi contact ($\tilde{A}_{iso}$) and dipolar ($\tilde{A}_{dip}$)
$^{13}$C HFCCs vs the corresponding values, 
$A_{iso}$ (b) and $A_{dip}$ (c), calculated from first principles. 
The values for inner (3 carbon NNs) and boundary (2 carbon NNs) atoms are shown
as red and blue dots, respectively.}
\end{figure}     

\begin{table}
\caption{\label{tab1} Parameters (in~MHz) fitted to the results
of calculations of a set of nanographite molecules.}
\begin{ruledtabular}
\begin{tabular}{ccccccc}
   &  $a_2$ & $a_3$ & $b_2$ & $b_3$ & $c$ & $d$ \\
\hline
 $A_{iso}$ & 162   &  128   & 2.3   & 6.1   &  -57.4   & -7.4 \\
 $A_{dip}$ & 155     & 131    & 2.7    &  3.6    & -19.4       & -12.8   
\end{tabular}
\end{ruledtabular}
\end{table}      

The HFIs were calculated for a set of 12 ($\sim$1~nm size) electron- and hole-doped
planar hydrogen terminated graphene fragments (Fig.~\ref{fig2}a) 
in the spin-doublet ground states.
This provides overall statistics for 206 inequivalent $^{13}$C HFCCs.  
The calculated $A_{iso}$ and $A_{dip}$ values are fitted to the extended form of the Karplus-Fraenkel expression 
\begin{eqnarray}
\label{eq1}
	A = a_j (1 + \sum_{i\in NN} b_j \Delta r_i)n^c + c \sum_{i\in NN}(1 + d\Delta r_i)n^c_i,
\end{eqnarray}
where the two terms account for the contributions of the on-site and 
nearest neighbor (NN) conduction electron spin populations per unpaired electron, $n^c$ and 
$n^c_i$, respectively, calculated from first principles.
The on-site coefficients
$a_j$ and $b_j$ are distinguished for the cases of C atoms with 3 carbon 
NNs ($j$=3) and the boundary atoms with 2 carbon NNs ($j$=2).
The C--C bond length effects are encountered through the coefficients 
$b_j$ and $d$ with $\Delta r_i$$=$$r_i$$-$$r_0$ being the deviation of the bond 
length $r_i$ from the value for the ideal graphene, $r_0$=1.42~\AA. 
Only statistically significant local properties were included in the linear expression (\ref{eq1}).
The results of the regressions are summarized in Tab.~\ref{tab1} (1~MHz=4.136$\times$10$^{-3}$~$\mu$eV). 
Fig.~\ref{fig2}(b,c) shows the fitted (using expr.~(\ref{eq1}) and regression parameters) 
values $\tilde{A}_{iso}$ ($\tilde{A}_{dip}$) versus the 
calculated $A_{iso}$ ($A_{dip}$) values. Regressions to the linear expression (\ref{eq1}) 
provide accurate estimations 
(root-mean-square-errors are 1.7~MHz and 1.2~MHz for $A_{iso}$ and $A_{dip}$, respectively).
The calculated isotropic HFCCs
span about the same range of magnitudes ($-$16.7~MHz$<$$A_{iso}$$<$21.1~MHz) as
the dipolar HFCCs ($-$5.9~MHz$<$$A_{dip}$$<$22.7~MHz).
The HFCCs of boundary atoms tend to be larger due to the fact that 
low-energy states localize at the zigzag graphene edges \cite{Nakada96}. 
The on-site and the NN exchange-polarization effects have competitive character
($a_3$/$c$$\approx$$-$2) in the case of isotropic HFCCs. 
Our calculations predict $\approx$50\% 
larger values for the parameters $a_2$, $a_3$ and  $c$ for $A_{iso}$ compared to those 
obtained by Karplus and Fraenkel in their early studies of HFCCs in molecular radicals
($a_2$=$99.8$~MHz, $a_3$=$85.5$~MHz and $c$=$-$39~MHz) \cite{Karplus61}.
This difference can be explained by the incorporation (via DFT) of the electron 
correlation effects in our calculations and to the 
local atomic structure of the graphene lattice.
Both $A_{iso}$ and $A_{dip}$ show a tendency to enhance the on-site 
and to weaken the NN contributions with the increase of C--C bond lengths.
The dipolar HFCC is mostly influenced
by the on-site contribution of the half-populated conduction state and
the NN exchange-polarization effect is weaker in this case ($a_3$/$c$$\approx$$-$7). 
When compared to typical solid state environments
based on heavier elements,
the $^{13}$C HFCCs in graphene and related nanostructures are weaker (e.g. 117~MHz $^{31}$P
Fermi contact HFCC for the P shallow donor in Si \cite{Overhof04}) and more anisotropic.

The graphene honeycomb lattice is a bipartite lattice, i.e. it can be 
partitioned into two complementary sublattices $A$ and $B$. 
I discuss the HFIs for the three general cases of 
conduction electron spin distributions over the sublattices:
(i) ferromagnetic $n^c_A$=$n^c_B$$>$$0$; (ii) ferrimagnetic 
$n^c_A$$>$$0$ and $n^c_B$=0 and (iii) antiferromagnetic $n^c_A$=$-n^c_B$$>$$0$
(see Tab.~\ref{tab2}).
The first case can be physically realized upon the uniform magnetization of 
the system with equivalent $A$ and $B$ sublattices, e.g. by applying an external magnetic field. 
The negative $A_{iso}$$=$$-$44~MHz is  small 
due to the partial compensation of the on-site and the NN 
exchange-polarization effects. This value is consistent
with the values derived from the experimental $^{13}$C Knight shifts in
graphite intercalates ($-$25~MHz$<$$A_{iso}$$<$$-$50~MHz) \cite{Conard80} and 
with the calculated isotropic Knight shifts in metallic carbon 
nanotubes \cite{Yazyev05}. 
The ferrimagnetic case with the conduction state distributed over 
the atoms of only one sublattice ($A$) is physically realized at the zigzag edges 
\cite{Nakada96} and around 
single-atom point defects in sublattice $B$ \cite{Kelly98}. 
Considerable alternating Fermi contact and dipolar 
HFCCs are predicted in this case. An antiferromagnetic pattern can be realized
in the case of heavily disordered systems with localized defect and edge states in both sublattices \cite{Yazyev06}. 
The magnitudes of HFIs are minimized and maximized in the cases of ferromagnetic and antiferromagnetic
electron spin distributions, respectively. 

\begin{table}
\caption{\label{tab2} Hyperfine coupling constants 
for three general cases of spin populations $n^c$ of the carbon atoms 
of $A$ and $B$ sublattices of graphene.}
\begin{ruledtabular}
\begin{tabular}{lcccc}
   & $A_{iso}$(A)/$n^c$  &  $A_{iso}$(B)/$n^c$  & $A_{dip}$(A)/$n^c$  &  $A_{dip}$(B)/$n^c$  \\
\hline
  $n^c_A$=$n^c_B$$>$$0$     &   -44    &    -44       &    73   &   73   \\
  $n^c_A$$>$$0$; $n^c_B$=0  &   128    &    -172      &    131  &  -58   \\
   $n^c_A$=$-n^c_B$$>$$0$   &   300    &    -300      &    189  &  -189
\end{tabular}
\end{ruledtabular}
\end{table}

Many carbon nanostructures of reduced dimensionality (e.g. nanotubes and fullerenes) 
represent non-planar topologies. Local curvatures lead to the 
$sp^2$$-$$sp^3$ rehybridization of carbon atoms and enable a Fermi contact 
interaction involving the low-energy $\pi$ electron spins \cite{Pennington96}. 
This results in a positive contribution of the $\pi$ 
states unless $n^c$ is close to zero: a contribution due to the NN
exchange-polarization effect is negative in this case.
The degree of rehybridization $m$ of the $\pi$ states ($s^m p$) can be described
using a local bond angles analysis \cite{Haddon86}.
For the case of large curvature radii the original expression for $m$ can be reformulated 
in a more convenient form,
$m$=$d_{cc}^2/8(1/R_1 + 1/R_2)^2$,
where $d_{cc}$ is the C--C distance, $R_1$ and $R_2$ are the principal 
curvature radii. 
The curvature-induced contribution to the Fermi contact 
$^{13}$C HFCC is then $A_{iso}^{curv}$$=$$(4\pi/3 S)\beta_e\beta_N g_e g_I n m \phi_{2s}^2(0)$, 
where $\phi_{2s}(0)$ is the magnitude of the carbon atomic
$2s$ wavefunction at the point of nucleus ($(8\pi/3)\beta_e\beta_N g_e g_I \phi_{2s}^2(0)$$\approx$3.5$\times$10$^3$~MHz). 
The curvature-induced direct coupling becomes significant ($m$$>$$10^{-2}$) only in ultranarrow  
carbon nanotubes ($d$$<$1~nm) and fullerenes.

\begin{table}
\caption{\label{tab3} Parameters (in~MHz) describing the HFIs of  
nuclei of susbtitutional 
impurities ($^{11}$B and $^{14}$N), monoatomic functional 
groups ($^{1}$H and $^{19}$F)
and rehybridized ($sp^3$) carbon atoms ($^{13}$C) at the edges. }
\begin{ruledtabular}
\begin{tabular}{cccccc}
 &    & \multicolumn{2}{c}{$A_{iso}$}  &  \multicolumn{2}{c}{$A_{dip}$}  \\
 Nucleus  & Position &   a  &  c  &  a  &  c  \\
\hline
  $^{11}$B  & subst. impurity  & 43 & -31 & 60 &  6 \\
  $^{14}$N  & subst. impurity  & 150 & -22 & 130 & -10 \\
  $^{1}$H   & C$_{sp^2}$ edge  & -119 & 22 & & \\
  $^{19}$F  & C$_{sp^2}$ edge  & 240 & -40 &  & \\
  $^{1}$H   & C$_{sp^3}$ edge  &   & 350 & & \\
  $^{19}$F  & C$_{sp^3}$ edge  &   & 750 & & \\
  $^{13}$C  & C$_{sp^3}$ edge  &   & -68 & & \\
\end{tabular}
\end{ruledtabular}
\end{table}

Since the natural abundance of the ``HFI-active'' $^{13}$C isotope is small ($\approx$1\%), 
consideration of the nuclei of other elements is important for a complete description
of HFIs	in carbon nanostructures. The common substitution impurities
are boron and nitrogen with all natural isotopes having nuclear spins. 
Graphene edges can be terminated by hydrogen and fluorine atoms with both 
$^1$H and $^{19}$F spin-1/2 nuclei (99.9885\% and 100\% natural abundance, 
respectively) having high $g$-values ($g(^{1}$H$)/g(^{13}$C$)$$\approx$$g(^{19}$F$)/g(^{13}$C$)$$\approx$4).
I consider HFIs in a reduced set of molecular fragments (only 3- and 4-ring structures included) with impurities and 
edge functionalizations in all possible positions. 
The calculated HFCCs have been fitted to the Karplus-Fraenkel
relation, with no $\Delta r$ terms included (Tab.~\ref{tab3}). 
The variations of the local charge density of states in the vicinity
of impurities does not have any significant influence on HFIs.
Both Fermi contact and dipolar HFCCs of the impurity nuclear spins 
show a monotonic increase along the $^{11}$B--$^{13}$C--$^{14}$N 
series when compared to
the results for $^{13}$C HFCCs (Tab.~\ref{tab1}). 
The NN relative exchange-polarization 
effects ($a/c$ ratio) on the Fermi contacts HFCCs tend to decrease along the series. 
While the HFIs of the nuclear spins in substitution impurities
are highly anisotropic, the hyperfine 
couplings of the edge nuclei show small anisotropy 
due to the $sp^3$ character of bonding. 
When $^1$H and $^{19}$F edge nuclei are bound to the C$_{sp^2}$
atoms, the isotropic HFCCs  are of the same order of 
magnitude as those of the $^{13}$C spins in the graphene lattice. The influence of the NN carbon atoms (second NNs
to the terminating atom) is very similar for $^1$H and $^{19}$F nuclei and smaller
than in the case of $^{13}$C HFCCs ($a_H/c_H$$\approx$$a_F/c_F$$\approx$$-$6). 
The spin polarization effect on
$^{19}$F HFCCs is stronger and of opposite sign compared to that of
protons ($a_F/a_H$$\approx$$c_F/c_H$$\approx$$-$2).
When edge atoms are bound to the 
rehybridized ($sp^3$) carbon atoms, $n^c$ is zero but the NN contribution 
is significantly enhanced. The NN contribution to the $^{13}$C 
hyperfine coupling of the $sp^3$ edge carbon atom itself ($c$=$-$68~MHz) 
has a similar magnitude as that of the $sp^2$ edge atoms ($c$=$-$57~MHz).  
HFIs with the boundary spins (H-terminated edges are often obtained in 
experiments \cite{Kobayashi06}) have to be taken into account when designing 
carbon-based nanoscale devices for spintronics or quantum computing. 
A chemical modification of the graphene edges (e.g. substitution of the hydrogen atoms by
alkyl-groups) can be suggested to reduce electron spin decoherence effects from the HFIs
with boundary spins.  

In conclusion, the results of first principles calculations show
that the hyperfine interactions in graphene and related nanostructures are
defined by the local distribution of the conduction electron spins
and by the local atomic structure.
A complete set of parameters describing the hyperfine interactions  
was determined for the $^{13}$C and other common nuclear spins. 
These results will permit control of the magnetic interactions 
between the spins of electrons and nuclei by tailoring the chemical 
and isotopic compositions, local atomic structures, and 
strain fields in $sp^2$ carbon nanostructures. Some practical recipes for minimizing
interactions with nuclear spins are given.

I acknowledge D.~Loss and Yu. G. Semenov for motivating discussions, and 
S.~Arey, D.~Bulaev, L.~Helm, V.~G.~Malkin, D.~Stepanenko, and I.~Tavernelli for 
comments on the manuscript.
I also thank the Swiss NSF for financial support and 
CSCS Manno for computer time.

\end{document}